\documentclass[aps,prl,preprint,superscriptaddress]{revtex4-2}

\usepackage{graphicx}
\usepackage{dcolumn}
\usepackage{bm}
\usepackage[colorlinks=true,allcolors=blue]{hyperref}

\usepackage{color}

\begin{document}
	
	\title{Influence of sulphur vacancies on ultrafast charge separation in WS$_2$-graphene heterostructures}
	
	
	
	\author{Johannes Gradl}
	\affiliation{Department for Experimental and Applied Physics, University of Regensburg, Regensburg, Germany}
	
	\author{Niklas Hofmann}
	\affiliation{Department for Experimental and Applied Physics, University of Regensburg, Regensburg, Germany}
	\author{Leonard Weigl}
	\affiliation{Department for Experimental and Applied Physics, University of Regensburg, Regensburg, Germany}
	\author{Stiven Forti}
		\affiliation{Center for Nanotechnology Innovation@NEST, Istituto Italiano di Tecnologia, Pisa, Italy}
	\author{Neeraj Mishra}
		\affiliation{Center for Nanotechnology Innovation@NEST, Istituto Italiano di Tecnologia, Pisa, Italy}
		\affiliation{Graphene Labs, Istituto Italiano di Tecnologia, Genova, Italy}
	\author{Camilla Coletti}
		\affiliation{Center for Nanotechnology Innovation@NEST, Istituto Italiano di Tecnologia, Pisa, Italy}
	\author{Raul Perea-Causin}
		\affiliation{Department of Physics, Stockholm University, AlbaNova University Center, Stockholm, Sweden}	
	\author{Ermin Malic}
		\affiliation{Department of Physics, University of Marburg, Marburg, Germany}
	\author{Isabella Gierz}
	\email[]{isabella.gierz@ur.de}
		\affiliation{Department for Experimental and Applied Physics, University of Regensburg, Regensburg, Germany}
	
	
	\date{\today}
	
	\begin{abstract}
	
Understanding how defects influence charge separation in WS$_2$-graphene heterostructures is crucial for future applications in light harvesting and detection. Previous studies have reported widely varying lifetimes for the charge-separated state, all supposedly linked to electron trapping at sulphur vacancies. The exact impact of these defects, however, has remained unclear. Here, we deliberately introduce sulphur vacancies by annealing the heterostructures at high temperatures in ultrahigh vacuum. Angle-resolved photoemission spectroscopy (ARPES) reveals that these vacancies modify both the band alignment and doping level of the heterostructure. Time-resolved ARPES (trARPES) further shows that increasing the sulphur vacancy concentration prolongs the lifetime of electrons in the WS$_2$ conduction band but shortens the lifetime of the charge-separated state. Guided by model calculations, we attribute this behaviour to shifts in the energy alignment between sulphur vacancy states and graphene’s Dirac point, combined with a reduced excitonic absorption. The model also yields a transfer time for electrons tunneling from sulphur vacancies into graphene’s Dirac cone of $\sim$4\,ps, consistent with our trARPES measurements. Our study clarifies the role of sulphur vacancies in WS$_2$-graphene heterostructures, further improving our microscopic understanding of charge dynamics for future optoelectronic applications.
	
	\end{abstract}
	
	
	\maketitle

\section{Introduction}
	
WS$_2$-graphene heterostructures show efficient ultrafast charge separation following photoexcitation at resonance to the A-exciton in monolayer WS$_2$ \cite{Yuan_SciAdv2018, Aeschlimann_SciAdv2020, Krause_PRL2021, Fu_SciAdv2021, Trovatello_NPJ2022}. The lifetime of the charge-separated state has remained highly controversial with numbers ranging from $\sim$1\,ps \cite{Aeschlimann_SciAdv2020, Krause_PRL2021} to $\sim$1\,ns \cite{Fu_SciAdv2021, Fu_NanoLett2023}. Despite differing by three orders of magnitude, these lifetimes have all been attributed to the trapping of electrons at sulphur vacancies within the WS$_2$ layer \cite{Krause_PRL2021, Fu_SciAdv2021, Fu_NanoLett2023}. As the lifetime of the charge-separated state determines the efficiency of possible light harvesting and detection devices, understanding the influence of defects on charge separation and recombination remains a crucial yet unresolved challenge.

Here, we deliberately create sulphur vacancies in WS$_2$-graphene heterostructures by annealing them at 650$^\circ$C in ultrahigh vacuum \cite{Liu_Small2017, Schuler_PRL2019, Mitterreiter_NatCommun2021} and investigate their influence on the band alignment with ARPES. We find that the presence of sulphur vacancies dopes the he\-te\-ro\-structure with additional electrons and reduces the energy separation between the top of the WS$_2$ valence band (VB) and the Dirac point of graphene. 

We also explore the influence of sulphur vacancies on ultrafast charge separation and recombination. For that purpose, we excite the heterostructure at resonance to the A-exciton in WS$_2$ and probe the relaxation of the photoexcited electron-hole pairs with trARPES. We find that the lifetime of the photoexcited electrons inside the WS$_2$ conduction band (CB) increases with increasing sulphur vacancy concentration, $n_v$, while the lifetime of the charge-separated state decreases with increasing $n_v$. We attribute these findings to two effects: First, the observed change in band alignment increases the energy offset between sulphur vacancy levels and the Dirac point of graphene and thereby the tunneling rate of electrons from sulphur vacancies into the Dirac cone. Second, recent GW-BSE calculations \cite{Hernangomez_NanoLett2023} predict that the presence of sulphur vacancies leads to a considerable reduction of the excitonic absorption in WS$_2$-graphene heterostructures which is expected to increase the lifetime of electrons inside the WS$_2$ CB and decrease the lifetime of the charge-separated state \cite{Krause_PRL2021}.

To quantify electron transfer from sulphur vacancy states in WS$_2$ to the Dirac cone of graphene, we model the time evolution of the electron density using a second-order Born-Markov approach parametrized by density functional theory \cite{Hernangomez_PRB2023}. The model yields a transfer time of $\sim$4\,ps in good agreement with our present and previously published results \cite{Aeschlimann_SciAdv2020, Krause_PRL2021, Hofmann_2DMater2023} and recent theoretical predictions \cite{Hernangomez_PRB2023}. This suggests that the $\sim$1\,ns lifetime observed by Fu et al. \cite{Fu_SciAdv2021, Fu_NanoLett2023} is either not due to sulphur vacancies or that the WS$_2$ layers used in these works were much thicker than one monolayer \cite{Bobzien_PRL2025}. 
	
\section{Methods}
WS$_2$-graphene heterostructures were grown on SiC(0001) substrates and characterized as previously described in \cite{Riedl_PRL2009, Emtsev_NatMat2009, Forti_Nanoscale2017, Fabbri_JPhysChemC2020}. WS$_2$ grows in the shape of triangular islands with an edge length of $\sim$ 500\,nm that are epitaxially aligned with the graphene lattice.

Following the procedure introduced in \cite{Liu_Small2017, Schuler_PRL2019, Mitterreiter_NatCommun2021} sulphur vacancies were deliberately induced by annealing the WS$_2$-graphene heterostructure in ultrahigh vacuum. We first performed four annealing steps at 600$^\circ$C for 30 minutes, each. Because neither the equilibrium band structure nor the carrier dynamics were affected by this, we subsequently increased the annealing temperature to 650$^\circ$C where we performed three additional annealing steps for 10, 30 and 30 minutes, respectively. The temperature during all annealing steps was monitored using an optical pyrometer (Pyrospot DG 10NV, DIAS Infrared Systems) with an emissivity of $\epsilon = 0.63$. 

For trARPES, we split the output of a 1\,kHz Titanium:Sapphire amplifier (Astrella, Coherent) into a pump and a probe arm. The pump arm was used to seed an optical parametric amplifier (Topas Twins, Light Conversion). The second harmonic of the signal output served as pump pulse with a photon energy of $\hbar \omega_{pump} = 2$\,eV and a pump fluence of 4\,mJ/cm$^2$. The second harmonic of the probe arm was used for high harmonic generation in an Argon gas jet. A single harmonic with a photon energy of $\hbar \omega_{probe} = 21.7$\,eV was selected with a grating monochromator and focused onto the sample with a toroidal mirror. The beam diameter at the focus was $\sim$300\,$\mu$m. Thus, the trARPES data represents an ensemble average over many different WS$_2$ islands. ARPES spectra were recorded with a hemispherical analyzer (Phoibos 100, Specs). The energy and temporal resolution of the present experiment were 260\,meV and 160\,fs, respectively.

\section{Results}

\begin{figure}
		\includegraphics[width = 1\columnwidth]{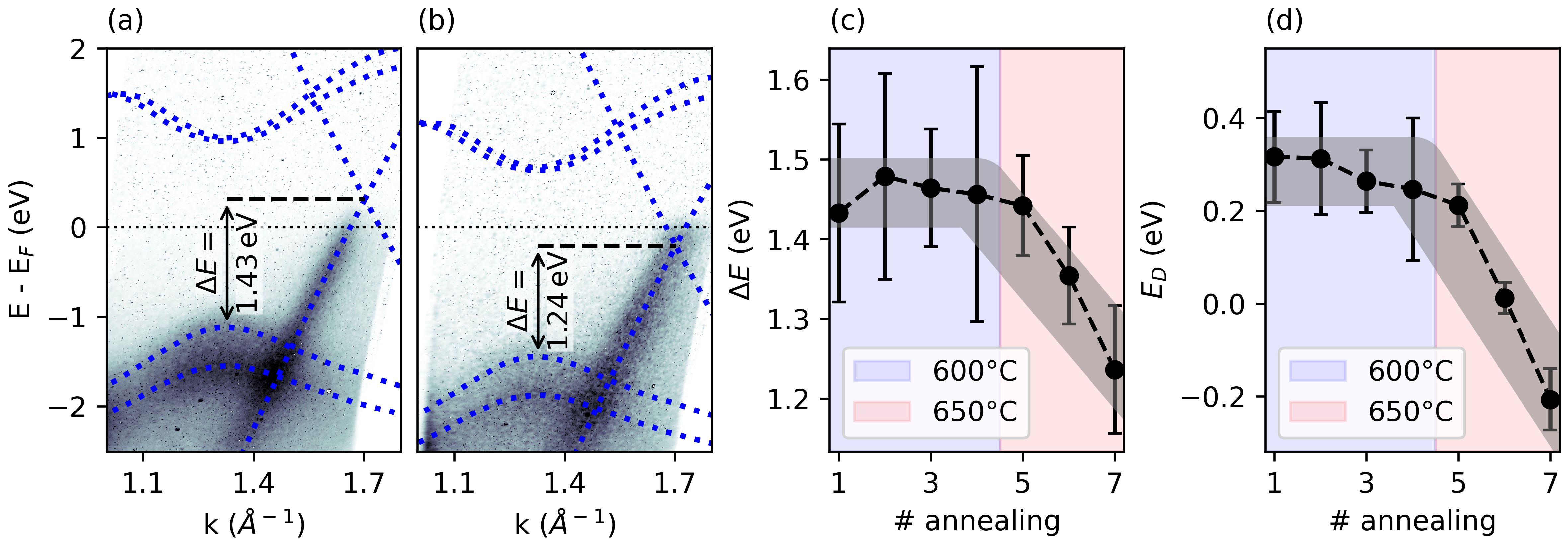}
		\caption{Influence of annealing on equilibrium band structure of WS$_2$-graphene \mbox{heterostructure}. ARPES snapshot of WS$_2$-graphene heterostructure after degassing the sample at 600$^\circ$C in UHV (a) and after four annealing steps at 600$^\circ$C and three annealing steps at 650$^\circ$C (b). Dashed blue lines in (a) and (b) are guides to the eye indicating the equilibrium band structure \cite{Zeng2013, Wallace1947}. (c) Distance between WS$_2$ VB maximum and Dirac point for different annealing steps. (d) Position of the Dirac point for different annealing steps. The blue- and red-shaded areas in (c) and (d) indicate annealing at 600$^\circ$C and at 650$^\circ$C, respectively. }
		\label{figure1}
	\end{figure}

In Fig.\,\ref{figure1} we present the effect of annealing on the equilibrium electronic structure of our WS$_2$-graphene heterostructure. Fig. 1a shows an ARPES spectrum measured close to the $K$-point along the $\mathit{\Gamma K}$ direction of the hexagonal Brillouin zone after degassing the sample at 600$^\circ$C in ultrahigh vacuum. The spectrum clearly reveals the spin-orbit split VB of WS$_2$ and the linear dispersion of the Dirac cone of graphene. The distance between the maximum of the WS$_2$ VB and the Dirac point is found to be $\Delta$E $\sim$ 1.43\,eV. The Dirac cone is found to be hole doped with the Fermi level $\sim$320\,meV above the Dirac point. Repeated annealing of the sample at 600$^\circ$C (annealing steps 1-4) had no significant influence on the band structure and carrier dynamics (see Fig.\,\ref{figure2}) which is why we increased the annealing temperature to 650$^\circ$C (annealing steps 5-7). The snapshot in Fig.\,\ref{figure1}b represents the ARPES spectrum taken after the final annealing step. We find that the distance between the maximum of the WS$_2$ VB and the Dirac point decreased to $\sim$1.24\,eV and the Dirac point of graphene dropped below the Fermi level (E$_{\text{D}}$ $\sim$ $-$210\,meV). The annealing-induced changes of the band structure are summarized in Figs.\,\ref{figure1}c and d. We find that annealing at 650$^\circ$C results in a decrease of the distance between the maximum of the WS$_2$ VB and the Dirac point, $\Delta E$, (Fig.\,\ref{figure1}c) and an increase of the binding energy of the Dirac cone, $E_{D}$ (Fig.\,\ref{figure1}d).

References \cite{Liu_Small2017, Schuler_PRL2019, Mitterreiter_NatCommun2021} showed that vacuum annealing at temperatures $\geq$ 600$^\circ$C results in the desorption of sulphur atoms and the generation of single sulphur vacancies. Reference \cite{Liu_Small2017} further showed that the presence of sulphur vacancies in monolayer MoS$_2$ leads to n-doping. Therefore, we attribute the observed changes of the band alignment and doping level induced by annealing at 650$^\circ$C to the successful generation of sulphur vacancies. Note that H desorption from the G/H-SiC interface is known to set in at $>700^\circ$C and to reduce the p-doping of the Dirac cone \cite{Sieber_PRB2003, Riedl_PRL2009}. While our maximum annealing temperature was $50^\circ$C below this threshold, we cannot entirely exclude the possibility that some of the band structure changes in Fig.\,\ref{figure1} result from H desorption.

In Fig.\,\ref{figure2} we present the influence of sulphur vacancies on the non-equilibrium carrier dynamics of the WS$_2$-graphene heterostructure. Figure \ref{figure2}a shows the pump-induced changes of the photocurrent of the ARPES snapshot in Fig.\,\ref{figure1}a at a pump-probe delay of $\sim$100\,fs after photoexcitation at $\hbar \omega_{pump} = 2$\,eV with a fluence of 4\,mJ/cm$^2$. To analyze this data in more detail, we integrate the photocurrent over the areas marked by the boxes in Fig.\,\ref{figure2}a and fit the resulting pump-probe traces with an exponential decay yielding the lifetime $\tau$. Further details are provided in Appendix I. We find that the lifetime of the carriers inside the WS$_2$ CB (orange box in Fig.\,\ref{figure2}a) increases with each annealing step at 650$^\circ$C (see Fig.\,\ref{figure2}b) and that the lifetime of the gain above the equilibrium position of the WS$_2$ VB (black box in Fig.\,\ref{figure2}a) decreases with each annealing step at 650$^\circ$C (see Fig.\,\ref{figure2}c).

In the transient charge-separated state, the graphene layer is positively charged and the WS$_2$ layer is negatively charged. This has been shown to lead to a transient down-shift of the graphene bands and a transient up-shift of the WS$_2$ bands \cite{Aeschlimann_SciAdv2020, Krause_PRL2021, Hofmann_2DMater2023}. Further, the WS$_2$ band gap was found to shrink due to the presence of photoexcited electron-hole pairs that screen the Coulomb interaction \cite{Ugeda_NatMater2014, Chernikov_NaturePhoton2015,Ulstrup_ACSNano2016, Liu_PRL2019, Krause_PRL2021, Hofmann_2DMater2023}. Thus, the gain signal above the equilibrium position of the WS$_2$ VB contains contributions from a transient up-shift due to band gap renormalization and layer charging as well as a transient broadening of the bands. We extracted the charging contribution to the up-shift of the VB as described in Appendix II and determined the exponential lifetime of the charging shift for different annealing steps from exponential fits. The result is shown in Fig.\,\ref{figure2}d. We find that the lifetime of the charge-separated state decreases with each annealing step at 650$^\circ$C.
	
In summary, Figs.\,\ref{figure2}c and d confirm our previously reported lifetime of the charge-separated state of some picoseconds \cite{Aeschlimann_SciAdv2020, Krause_PRL2021, Hofmann_2DMater2023}. In addition, Fig.\,\ref{figure2}b directly shows that the lifetime of photoexcited electrons inside the WS$_2$ CB gets longer with increasing $n_v$. Further, Figs.\,\ref{figure2}c and d show that the lifetime of the charge-separated state decreases with increasing $n_v$.

\section{Discussion}

To interpret the results obtained above we start by summarizing our previously developed microscopic model for ultrafast charge transfer in WS$_2$-graphene heterostructures \cite{Aeschlimann_SciAdv2020, Krause_PRL2021, Hofmann_2DMater2023} with the help of the band structure sketched in Fig.\,\ref{figure3}. Despite the weak van der Waals interaction between WS$_2$ and graphene the respective bands are found to hybridize in those regions of the Brillouin zone that are highlighted in red in Fig.\,\ref{figure3} \cite{Hofmann_2DMater2023}. In these areas, the corresponding wave functions are delocalized over both the graphene and the WS$_2$ layer, enabling efficient charge transfer between the layers. Electrons (holes) in the $K$ valley of the WS$_2$ CB (VB) need to overcome an energy barrier $\Delta E_{CB}$ ($\Delta E_{VB}$) in order to reach the closest charge transfer state. Electrons and holes then relax to the Fermi level located inside the Dirac cone of graphene by phonon emission. The lifetime of holes in the WS$_2$ VB (limited by the temporal resolution of the experiment) was found to be much shorter than the lifetime of electrons inside the WS$_2$ CB ($\sim$1\,ps) \cite{Aeschlimann_SciAdv2020, Krause_PRL2021}. This was explained by a smaller tunneling barrier, a bigger tunneling matrix element and a bigger density of available final states for holes compared to electrons \cite{Krause_PRL2021}. 	
	
In addition to this direct tunneling channel, electrons inside the WS$_2$ CB can be trapped by two unoccupied spin-orbit-split mid-gap states originating from sulphur vacancies \cite{Schuler_PRL2019} (yellow in Fig.\,\ref{figure3}).	These mid-gap states also hybridize with the graphene Dirac cone \cite{Hernangomez_PRB2023} opening a parallel defect-mediated charge transfer channel for electrons. The relative importance of these two parallel charge transfer channels depends on the electronic temperature (the higher the temperature, the easier it is for the electrons and holes to overcome the tunneling barrier) and $n_v$ (the higher $n_v$, the higher the probability for an electron to transfer into the graphene layer through a sulphur vacancy). In our previous work \cite{Krause_PRL2021} we found that direct tunneling sets the timescale for charge separation while defect-mediated tunneling sets the timescale for charge recombination. 

To shed light on the controversial lifetime of the charge-separated state in WS$_2$-graphene heterostructures with values ranging from $\sim$1\,ps in our present and previous work \cite{Aeschlimann_SciAdv2020, Krause_PRL2021, Hofmann_2DMater2023} to $\sim$1\,ns reported in the literature \cite{Fu_SciAdv2021, Fu_NanoLett2023}, we estimate the transfer time for an electron to tunnel from a sulphur vacancy in WS$_2$ into the Dirac cone of graphene in second-order Born-Markov approximation \cite{Malic_Knorr_2013}. Density functional theory calculations \cite{Hernangomez_PRB2023} predict that the lower vacancy level, located at an energy of $E_0\sim0.3$\,eV above the Dirac point, shows stronger hybridization with the Dirac cone than the one at higher energy (see sketch in Fig.\,\ref{figure3}). Therefore, we expect tunneling to occur mainly through this lower vacancy level where the size of the avoided crossing between sulphur vacancy and Dirac cone is $\delta E\sim30$\,meV for a vacancy concentration of $n_v=1\ \text{nm}^{-2}$ \cite{Hernangomez_PRB2023}. The temporal evolution of the electron density $n$ trapped at sulphur vacancies is given by $\dot{n}=-\frac{2\pi}{\hbar A}\sum_\mathbf{k} |t_\mathbf{k}|^2 \rho_\mathbf{k}\delta(\hbar v|\mathbf{k}|-E_0)$, where $A$ is the area of the unit cell, $t_\mathbf{k}$ the tunneling strength, $\rho_\mathbf{k}$ the probability that the vacancy state is occupied and $v=1$\,nm/fs the Fermi velocity in graphene. Evaluating this expression as $\dot{n}=-\tau^{-1}n$ and using $n=n_v\rho$, we obtain the transfer time $\tau = \hbar^3v^2n_v/(|t|^2E_0)$ with $t=1/2 \cdot \delta E\propto\sqrt{n_v}$ such that $\tau$ is independent of $n_v$. We obtain a transfer time of $\tau\sim4$\,ps, in excellent agreement with the quantum master equation approach in \cite{Hernangomez_PRB2023} and our own experimental results in Figs.\,\ref{figure2}c und d and \cite{Aeschlimann_SciAdv2020, Krause_PRL2021, Hofmann_2DMater2023}. Therefore, the $\sim$1\,ns lifetime of the charge-separated state observed in \cite{Fu_SciAdv2021, Fu_NanoLett2023} cannot be easily reconciled with trapping of electrons at sulphur vacancies in monolayer WS$_2$. We speculate that the commercial WS$_2$-graphene samples used in \cite{Fu_SciAdv2021} and the WS$_2$-graphene device used in \cite{Fu_NanoLett2023} may contain other defects and impurities or thicker WS$_2$ layers that might be responsible for the extremely long lifetimes of the charge-separated state \cite{Bobzien_PRL2025}.

Next, we discuss two possible scenarios that might in principle account for our observations in Figs.\,\ref{figure2}b-d. (1) The observed changes in band alignment (see Fig.\,\ref{figure1}) affect the tunneling barriers for direct tunneling $\Delta E_{VB}$ and $\Delta E_{CB}$ as well as the energy separation between sulphur vacancy states and the Dirac point $E_0$ for tunneling through sulphur vacancies. (2) Reference \cite{Hernangomez_NanoLett2023} predicted that the presence of sulphur vacancies results in a reduced absorption at the A-exciton resonance in WS$_2$-graphene heterostructures. 

(1) Figure \ref{figure1}c clearly shows that the presence of sulphur vacancies affects the band alignment and therefore the height of the tunneling barriers for electrons and holes. Based on Fig.\,\ref{figure1}c and Fig.\,\ref{figure3} and assuming a rigid band shift, $\Delta E_{CB}$ is expected to decrease and $\Delta E_{VB}$ is expected to increase with increasing $n_v$. Therefore, the rate for electron transfer is expected to increase and the rate for hole transfer is expected to decrease. As hole transfer happens on a timescale that is fast compared to the temporal resolution of our experiment, we cannot assess the influence of sulphur vacancies on the transfer rate of holes. On the other hand, Fig.\,\ref{figure2}b shows that the lifetime of photoexcited electrons in the WS$_2$ CB increases with increasing $n_v$, excluding changes in $\Delta E_{CB}$ as a possible explanation for our observations. However, in our model, the tunneling time for electrons through sulphur vacancies, $\tau$, is inversely proportional to the energy separation between sulphur vacancy levels and the Dirac point, $E_0$. Assuming rigid band shifts, the decrease of $\Delta E$ in Fig.\,\ref{figure1}c directly translates into an increase of $E_0$ from $\sim0.3$\,eV to $\sim0.5$\,eV which is expected to result in a corresponding reduction of $\tau$ in agreement with our observations in Figs.\,\ref{figure2}c and d.

(2) All our experiments were performed with the same pump fluence of 4\,mJ/cm$^2$. A reduced absorption due to sulphur vacancies would result in smaller electron-hole pair densities and lower carrier temperatures. Thus, increasing $n_v$ at constant pump fluence (Fig.\,\ref{figure2}) should have a similar effect on the charge carrier dynamics as reducing the pump fluence for a given $n_v$. The latter experiment was performed by us in the past \cite{Krause_PRL2021}. There, we showed that the lifetime of the carriers inside the WS$_2$ CB increases with decreasing fluence while the lifetime of the gain above the equilibrium position of the WS$_2$ VB decreases with decreasing fluence. Hence, our observations in Figs.\,\ref{figure2}b and \ref{figure2}c are also consistent with the scenario predicted in \cite{Hernangomez_NanoLett2023}.

Note that the relative change of $\tau_{charge}$  in Fig.\,\ref{figure2}d is a lot bigger than the one of $\tau_{CB}$ in Fig.\,\ref{figure2}b. Therefore, we speculate that the change of $\tau_{charge}$ is dominated by the increase of $E_0$, whereas the change of $\tau_{CB}$ is dominated by the reduced absorption.

\section{Summary}

We used trARPES to demonstrate that increasing the sulphur vacancy concentration in WS$_2$-graphene heterostructures leads to a longer electron lifetime in the WS$_2$ conduction band while shortening the lifetime of the charge-separated state. Supported by our own model as well as previously published theory \cite{Hernangomez_NanoLett2023} we attributed this to a combination of a change of the band alignment and reduced excitonic absorption. To address discrepancies in reported charge-separated state lifetimes, we modeled the electron transfer rate from sulphur vacancies in monolayer WS$_2$ to graphene, yielding a transfer time of $\sim$4\,ps. Future work should explore the role of different defect types and the layer thickness to refine the understanding of ultrafast charge separation in these heterostructures.

\section{Acknowledgments}

This work received funding from the European Union’s Horizon 2020 research and innovation program under Grant Agreement No. 851280-ERC-2019-STG as well as from the Deutsche Forschungsgemeinschaft (DFG) via the collaborative research centre CRC 1277 (Project No. 314695032), the collaborative research centre CRC 1083 (Project No. 223848855), the Research Unit RU 5242 (Project No. 449119662) and the regular project No. 512604469.

\clearpage

\pagebreak

	\begin{figure}
		\includegraphics[width = 0.5\columnwidth]{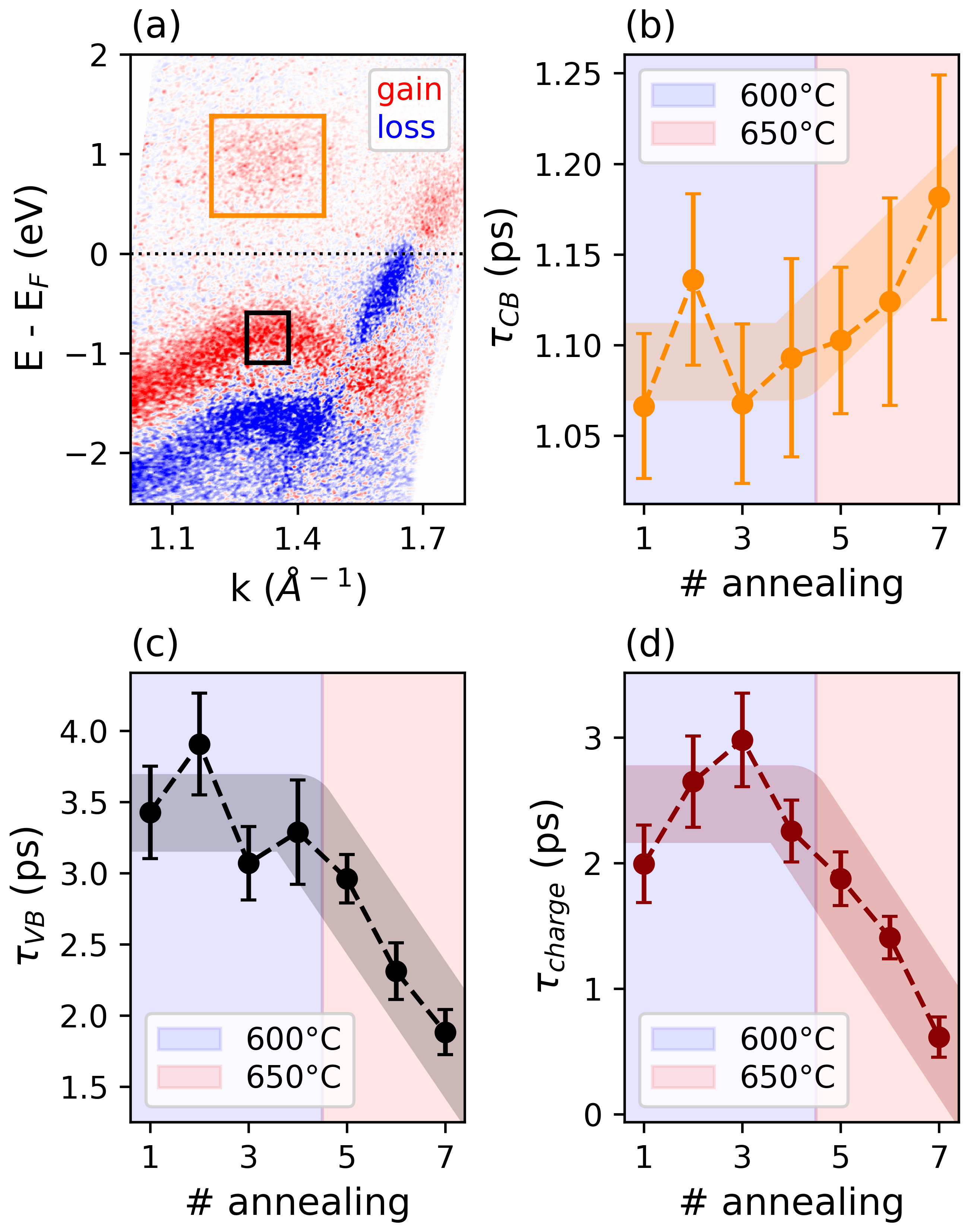}
		\caption{Influence of sulphur vacancies on non-equilibrium carrier dynamics. Panel a) shows the pump-induced changes of the photocurrent depicted in Fig. 1a for a pump-probe delay of $\sim$100\,fs after photoexcitation at $\hbar\omega_{pump} = 2$\,eV with a fluence of 4\,mJ/cm$^2$. b) Exponential lifetime of the photoexcited electrons in the WS$_2$ CB (yellow box in a) for each annealing step. c) Exponential lifetime of the photocurrent integrated over the area marked by the black box in (a) for each annealing step. The black box in (a) is consistently placed 25\,meV  above the equilibrium position of the WS$_2$ VB for all annealing steps. d) Exponential lifetime of the charging shift of the WS$_2$ VB for all annealing steps. Thick lines in (b-d) are guides to the eye.}
		\label{figure2}
	\end{figure}	
		
	\begin{figure}
		\includegraphics[width = 0.5\columnwidth]{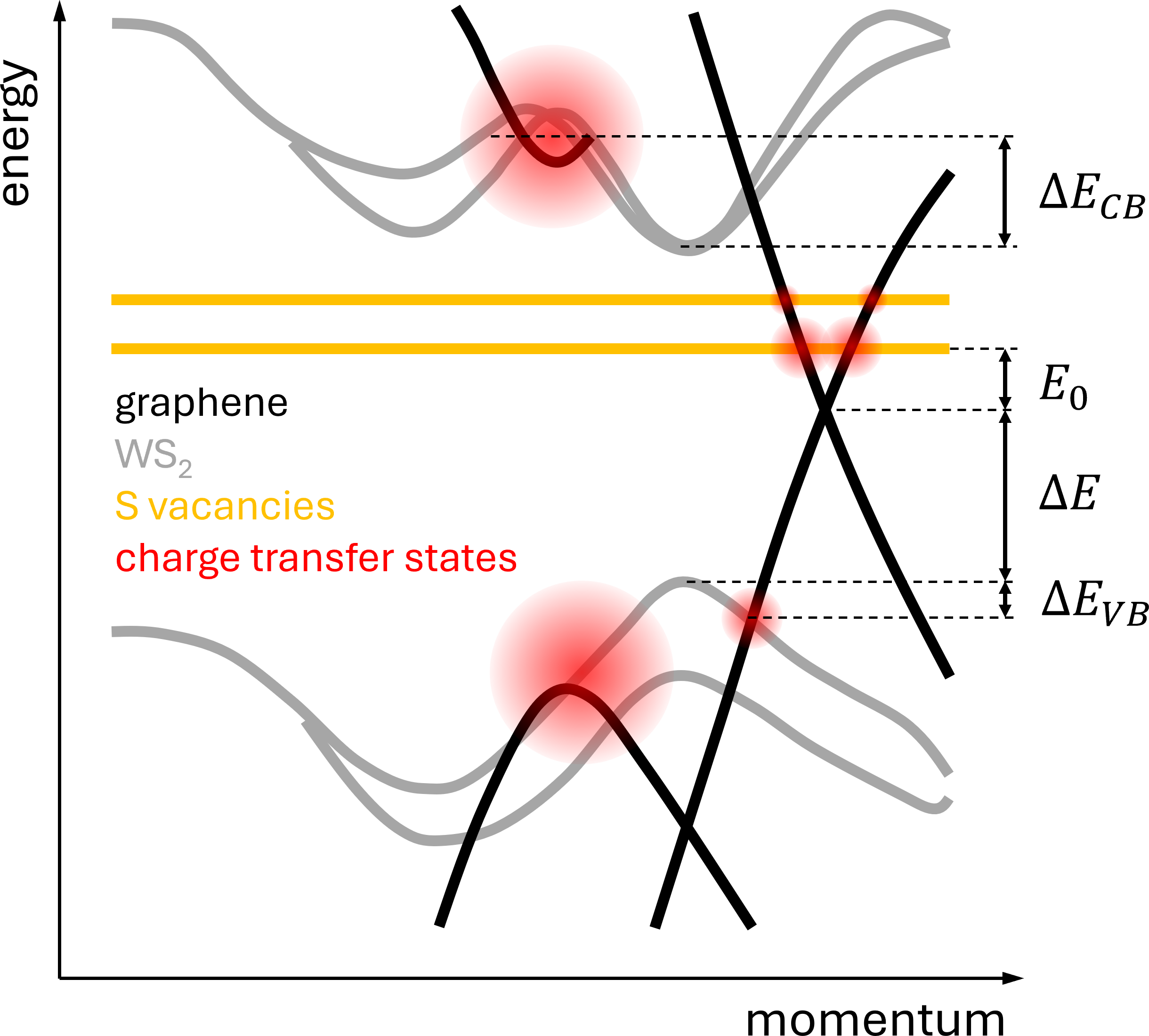}
		\caption{Sketch of band structure of WS$_2$-graphene heterostructure along the $\mathit{\Gamma K}$ direction based on DFT calculations from \cite{Hofmann_2DMater2023} and \cite{Hernangomez_PRB2023} to illustrate the microscopic model for charge transfer in WS$_2$-graphene developed in \cite{Krause_PRL2021}. Graphene (WS$_2$) bands are drawn in black (grey). Horizontal yellow lines indicate two mid-gap states originating from sulphur vacancies. Those areas in the Brillouin zone where the states hybridize and charge transfer can occur are highlighted in red. $\Delta E_{CB}$ and $\Delta E_{VB}$ indicate the barriers that electrons and holes need to overcome before they can tunnel from the WS$_2$ CB and VB into the $\pi$-bands of graphene. $\Delta E$ is the energy offset between the VB maximum and the Dirac point from Fig.\,\ref{figure1}c. $E_0$ is the distance between the lower sulphur vacancy level and the Dirac point that determines the charge transfer time from a sulphur vacancy to graphene.}
		\label{figure3}
	\end{figure}
	
\clearpage

\pagebreak

\section{Appendix I}

\begin{figure}[h]
		\includegraphics[width = 1\columnwidth]{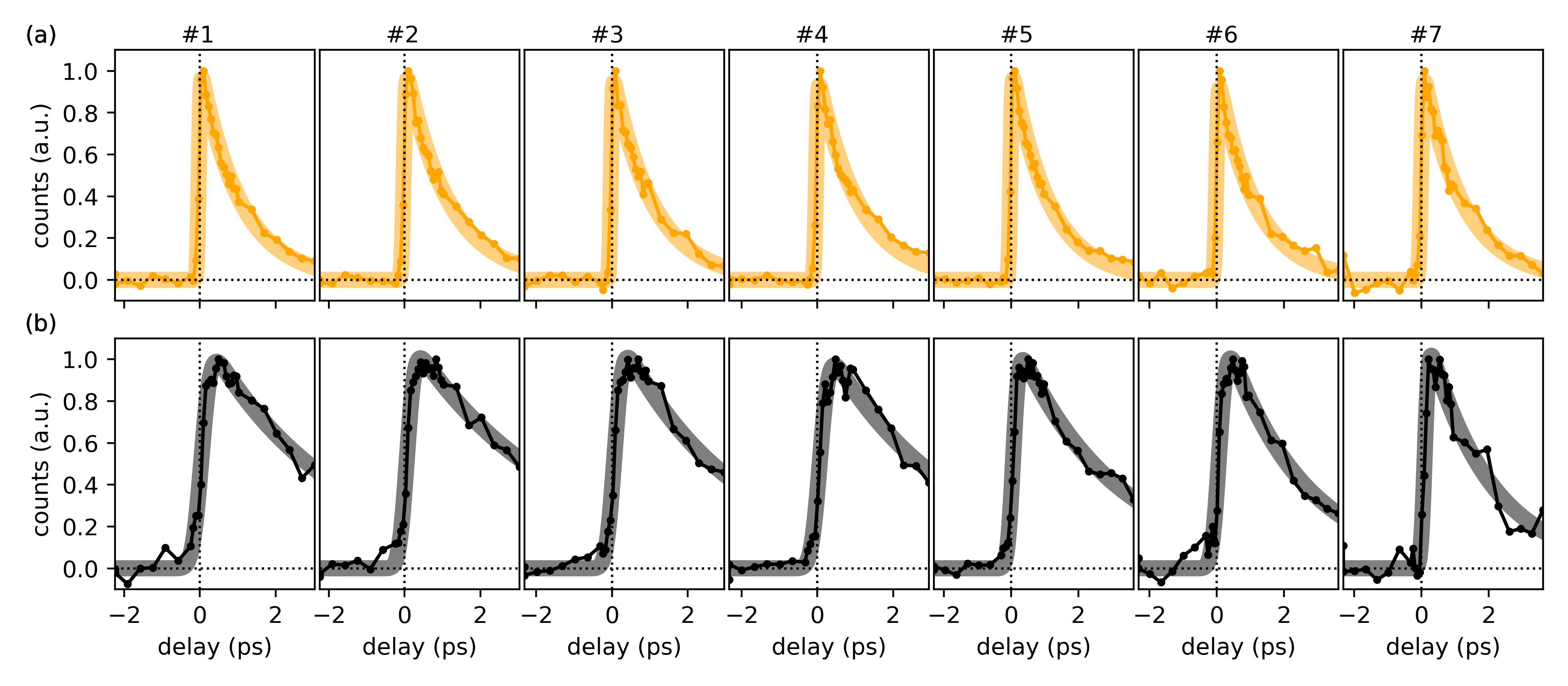}
		\caption{Pump-probe traces extracted by integrating trARPES snapshots from Fig.\,\ref{figure2}a over the areas marked by the yellow (a) and black boxes (b) together with exponential fits for all seven annealing steps.}
		\label{figureAppI}
	\end{figure}

In order to extract the lifetimes plotted in Fig.\,\ref{figure2}b we integrated the counts over the areas marked by the coloured boxes in Fig.\,\ref{figure2}a. The resulting pump-probe traces are shown in Fig.\,\ref{figureAppI} as a function of pump-probe delay for all seven annealing steps. The data points were fitted with a step function multiplied with an exponential decay. This product was then convolved with a Gaussian to account for the finite temporal resolution of the experiment. The resulting lifetimes are summarized in Table \ref{table}.

	\begin{table} [h]
\caption{Exponential lifetimes from Figs.\,\ref{figureAppI} and \ref{figureAppII} for different annealing steps. $\tau_{CB}$ and $\tau_{VB}$ are from Figs.\,\ref{figureAppI}a and b, respectively. $\tau_{charge}^{2.0\,eV}$ and $\tau_{charge}^{1.9\,eV}$ are from Figs.\,\ref{figureAppII}d and e, respectively.}
\label{table}
\begin{tabular}{ | c | c | c | c | c |}
\hline	
annealing step	& $\tau_{CB}$ (ps)	& $\tau_{VB}$ (ps)	& $\tau_{charge}^{2.0\,eV}$ (ps)	& $\tau_{charge}^{1.9\,eV}$ (ps) \\ 
\hline	
1											& 1.07$\pm$0.04		&  3.4$\pm$0.3				&  2.0$\pm$0.3													& \\ 
\hline	
2											& 1.14$\pm$0.05		&  3.9$\pm$0.4				&  2.7$\pm$0.4													& \\ 
\hline	
3											& 1.07$\pm$0.04		&  3.1$\pm$0.3				&  3.0$\pm$0.4													& \\ 
\hline	
4											& 1.09$\pm$0.05		&  3.3$\pm$0.4				&  2.3$\pm$0.3													& \\ 
\hline	
5											& 1.10$\pm$0.04		&  3.0$\pm$0.2				&  1.9$\pm$0.2													& 3.0$\pm$0.3 \\ 
\hline	
6											& 1.12$\pm$0.06		&  2.3$\pm$0.2				&  1.4$\pm$0.2													& 2.3$\pm$0.2 \\ 
\hline	
7											& 1.18$\pm$0.07		&  1.9$\pm$0.2				&  0.6$\pm$0.2													& 1.3$\pm$0.3 \\
\hline	 
\end{tabular}
\end{table}
	
\section{Appendix II}

\begin{figure}[h]
		\includegraphics[width = 1\columnwidth]{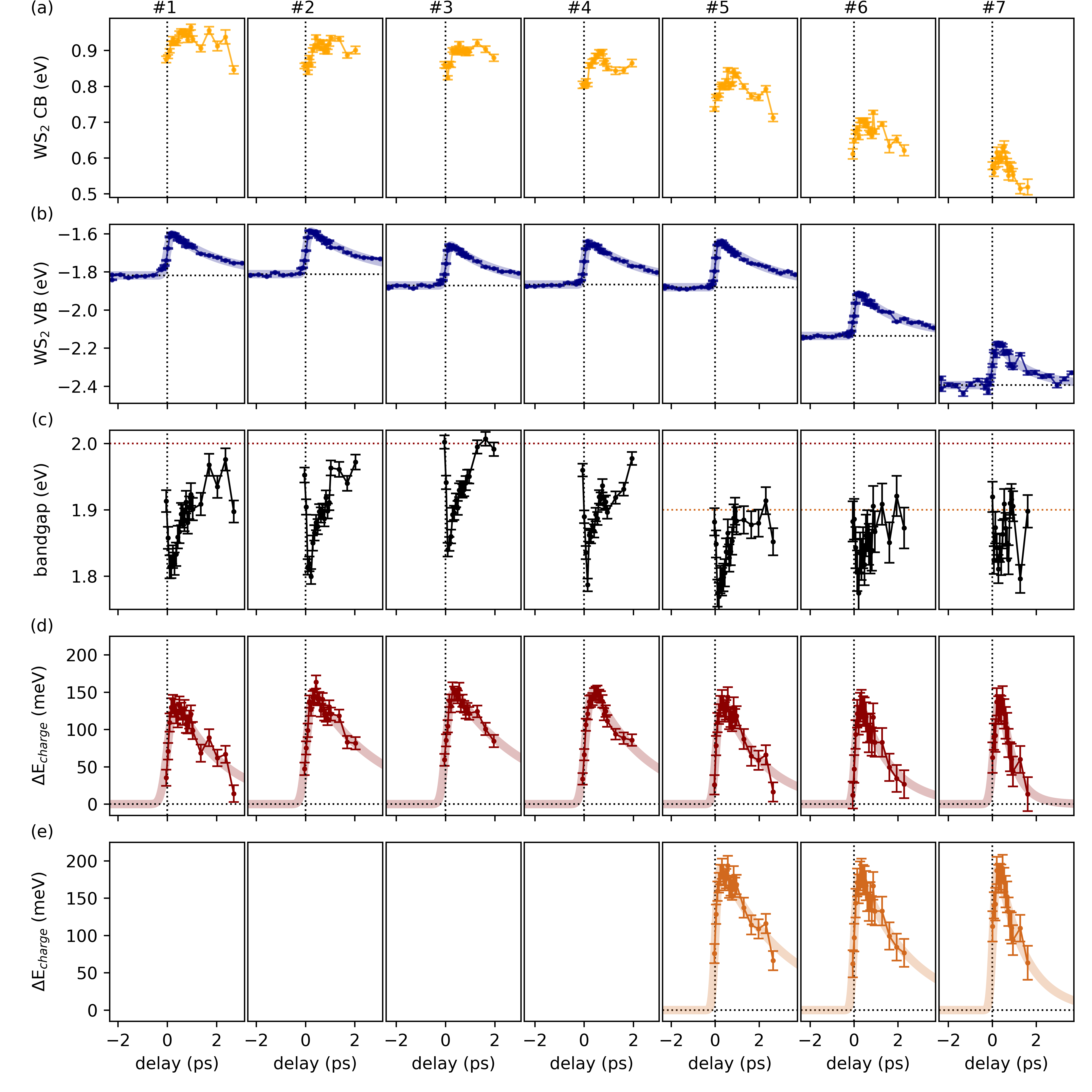}
		\caption{Band structure dynamics for all seven annealing steps as indicated in the individual panels. a) Transient position of the WS$_2$ CB at K. b) Transient position of the WS$_2$ VB at $k=1.1$\AA$^{-1}$. c) Transient band gap at $K$. d) Charging shift of the WS$_2$ VB under the assumption of a $2.0$\,eV equilibrium band gap together with exponential fit (continuous line). e) Same as d) but assuming an equilibrium gap size of 1.9\,eV.}
		\label{figureAppII}
	\end{figure}

In order to extract the lifetimes of the charging shift plotted in Fig.\,\ref{figure2}d we proceeded as follows. We extracted energy distribution curves at the $K$-point of WS$_2$ from Fig.\,\ref{figure2}a that we fitted with an appropriate number of Gaussian peaks and a Shirley background. These fits yielded the transient position of the WS$_2$ CB shown in Fig.\,\ref{figureAppII}a and the equilibrium position of the WS$_2$ VB. The transient position of the WS$_2$ VB was difficult to determine in this manner as the spin-orbit splitting could no longer be resolved due to the transient broadening of the VB. Instead, the transient position of the WS$_2$ VB shown in Fig.\,\ref{figureAppII}b was determined from similar EDC fits at $k=1.1$\AA$^{-1}$, where the VB can be described by a single Gaussian. Assuming that the transient up-shifts of the WS$_2$ VB at $K$ and $k=1.1$\AA$^{-1}$ are the same \cite{Hofmann_NanoLett2025}, we can then extract the transient band gap shown in Fig.\,\ref{figureAppII}c. Assuming an equilibrium band gap of 2.0\,eV and that the band gap changes symmetrically about its center, we can then subtract $\Delta E_{gap}/2$ from the total VB shift in Fig.\,\ref{figureAppII}b to obtain the charging shift of the WS$_2$ VB in Fig.\,\ref{figureAppII}d \cite{Krause_PRL2021}. This charging shift is then fitted with the same exponential function as described in Appendix I for the pump-probe traces. The resulting lifetimes are indicated in Table \ref{table}.

It has been reported that sulphur vacancies reduce the single-particle gap in monolayer MoS$_2$ \cite{Liu_Small2017}. Note that an accurate determination of the equilibrium band gap is difficult in the present case as --- at equilibrium --- the WS$_2$ CB is unoccupied. We would like to point out that our data in Fig.\,\ref{figureAppII}c for annealing steps 5-7 would also be consistent with a reduced equilibrium band gap of $\sim$1.9\,eV. To check whether this affects the trend observed in Fig.\,\ref{figure2}d, we repeated the data analysis for annealing steps 5-7 assuming an equilibrium band gap of 1.9\,eV (see Fig.\,\ref{figureAppII}e). Although the absolute numbers are found to increase (see Table \ref{table}), the trend of a decreased lifetime of the charge-separated state with increasing $n_v$ is robust. 

\clearpage

\pagebreak

\bibliography{literature}

	\end{document}